\documentstyle[12pt,aaspp4]{article}

\def \etal {{\it et al.}}

\lefthead{Bunker, Marleau \& Graham}
\righthead{Seeking the Ultraviolet Ionizing Background at $z\approx 3$ with the Keck Telescope}

\begin{document}

\title{Seeking the Ultraviolet Ionizing Background at $z\approx 3$ 
with the Keck Telescope}

\author{Andrew J.~Bunker\altaffilmark{1}, Francine R.~Marleau \& James
R.~Graham}
\affil{Astronomy Department, 601 Campbell Hall, University of
California, Berkeley CA~94720\\ {\tt email: bunker@bigz.Berkeley.EDU}}
\altaffiltext{1}{NICMOS Postdoctoral Researcher}

\begin{abstract}

We describe the initial results of a deep long-slit emission line search
for redshifted (2.7~$<z<$~4.1) Ly$\alpha$.  These observations are used
to constrain the fluorescent Ly$\alpha$ emission from the population of
clouds whose absorption produces the higher-column-density component of
the Ly$\alpha$ forest in quasar spectra. We use the results to set an
upper limit on the ultraviolet ionizing background. Our spectroscopic
data obtained with the Keck~II telescope at
${\lambda}/{\Delta\lambda_{\rm FWHM}}\approx$~2000 reveals no candidate
Ly$\alpha$ emission over the wavelength range of 4500--6200~\AA\ along a
3\arcmin\ slit in a 5400 s integration.  This null result places the
strongest limit to date on the ambient flux of Lyman continuum photons
at $z\approx$~3. Typically we attain a 1~$\sigma$ surface brightness
sensitivity to spectrally unresolved line emission in a 1~$\sq$\arcsec\
aperture of (0.4--0.6)$\times
10^{-18}$~erg~s$^{-1}$~cm$^{-2}$~arcsec$^{-2}$, and we search for
extended emission over a wide range of spatial scales. Our 3~$\sigma$
upper bound on the mean intensity of the ionizing background at the
Lyman limit is $J_{\nu0} < 2\times
10^{-21}$~erg~s$^{-1}$~cm$^{-2}$~Hz$^{-1}$~sr$^{-1}$ for 2.7~$<z<$~3.1
(where we are most sensitive), assuming Lyman limit systems have typical
radii of 70~kpc ($q_{0}=$~0.5, $H_0=$~50~km~s$^{-1}$~Mpc$^{-1}$).  This
constraint is more than an order of magnitude more stringent than any
previously published direct limit. However, it is still a factor of
three above the ultraviolet background level expected due to the
integrated light of known quasars at $z\approx$~3. This pilot study
confirms the conclusion of Gould \& Weinberg (1996) that integrations of
several hours on a 10-m class telescope should be capable of measuring
$J_{\nu0}$ at high redshift.  Our results suggest that the integrated
flux of Lyman continuum photons escaping from star-forming galaxies at
these epochs cannot exceed twice that from known quasars, and that the
completeness of optically-selected quasar catalogues must be better than
30\%. We also show that it is unlikely that decaying relic neutrinos, if
comprising the bulk of the dark matter, are responsible for the
meta-galactic radiation field.

\end{abstract}


\section{Introduction}

In most models of structure formation the majority of baryonic matter in
the Universe is contained in a uniformly-distributed intergalactic
medium (IGM), which is predicted as a product of primordial
nucleosynthesis.  The Gunn-Peterson constraint on the \ion{H}{1}
component of this IGM (Gunn \& Peterson 1965; Scheuer 1965) implies that
the gas must have been highly ionized by $z \sim$~5, because the flux
decrement observed on the blue side of the Ly$\alpha$ emission lines in
the spectra of high-redshift quasi-stellar objects (QSOs) appears to be
entirely consistent with absorption by discrete clouds along the line of
sight (Steidel \& Sargent 1987; Giallongo \etal\ 1994).

The meta-galactic ultraviolet (UV) background, which is widely believed
to be primarily the integrated light of QSOs, is thought to be
responsible for maintaining the diffuse IGM and the Ly$\alpha$ forest
clouds in a highly-ionized state.  This UV background may also be
responsible for the ionization of metal-rich QSO absorption systems
(Steidel 1990), the ionization of \ion{H}{1} clouds in the Galactic halo
(Ferrara \& Field 1994), as well as for producing the sharp edges of
\ion{H}{1} disks in nearby spirals (Dove \& Shull 1994; Maloney 1993;
Bland-Hawthorn, Freeman \& Quinn 1997).

From the evolution of the QSO luminosity function, recent theoretical
calculations for $z \approx$~2.5--3 indicate that the contribution of
observed QSOs alone to the mean intensity of the UV background is
$J_{\nu0(-21)}\approx$~0.6 at the Lyman limit, i.e., at a frequency
$\nu_0 = c~/~912$~\AA\ (Haardt \& Madau 1996, hereafter
HM96). Throughout we adopt the notation $J_{\nu0(-21)}=
J_{\nu0}~/~10^{-21}$~erg~s$^{-1}$~cm$^{-2}$~Hz$^{-1}$~sr$^{-1}$ and we
assume a cosmology of $q_0=$~0.5, $\Lambda_0=$~0 and
$h_{50}=H_0$~/~50~km~s$^{-1}$~Mpc$^{-1}$, unless otherwise stated.
Indirect estimates of $J_{\nu0}$ are uncertain (see
\S\ref{sec:discuss}) and, given its importance, strenuous efforts have
been made to directly measure the meta-galactic UV background.  Hogan
\& Weymann (1987) proposed that long-slit spectroscopy of ``blank
sky'' should reveal patches of fluorescent Ly$\alpha$ emission from
Ly$\alpha$ forest clouds excited by the meta-galactic background. A
detailed radiative-transfer treatment of Ly$\alpha$ emission from
thick clouds ($N_{\rm HI} > 10^{20}$~cm$^{-2}$) photo-ionized by the
meta-galactic radiation is given by Binette \etal\ (1993).

In a more recent paper, Gould \& Weinberg (1996) show that for clouds
with column density $N_{\rm HI} \gtrsim 3\times 10^{18}$~cm$^{-2}$
(optically thick to Lyman continuum photons), $\simeq$~50\% of the
incident energy of the ionizing background emerges in the form of
Ly$\alpha$ photons, including those resulting from collisional
excitation. This fraction is robust and independent of cloud geometry,
implying uniform surface brightness.  A Ly$\alpha$ photon is absorbed
and re-emitted within the cloud until it is Doppler shifted by an atom
with a velocity $\pm$~4~$\sigma_{v}$, at which point it can escape.  A
cloud with a typical one-dimensional (1D) velocity dispersion of
$\sigma_v =$~30~km~s$^{-1}$ would have a double-peaked Ly$\alpha$ line
with a separation of 8~$\sigma_{v}=$~240~km~s$^{-1}$. Hence,
moderate-resolution spectroscopy ($R={\lambda}/{\Delta\lambda_{\rm
FWHM}} \simeq$~1500) of an optically-thick cloud, along with adequate
integration time, will yield a direct measurement of the intensity of
the ionizing background.  Lowenthal \etal\ (1990) have observed the
Lyman limit system (LLS) towards Q0731$+$653, and on the basis of a
two hour exposure spectrogram with the Multiple Mirror Telescope
(MMT), place a 2~$\sigma$ upper limit of $J_{\nu0(-21)} <$~200 at $z
=$~2.91, on the assumption that the clouds are only 1\arcsec\ in
extent. Comparable sensitivity is obtained by
Mart\'{\i}nez-Gonz\'{a}lez \etal\ (1995) through narrow-band imaging
on the Isaac Newton Telescope (INT). Assuming larger clouds
($\sim$~10\arcsec), their 3~$\sigma$ upper-limit is equivalent to
$J_{\nu0(-21)} <$~80 at z~=~3.4 (see Figure~\ref{fig:jnulimits} and
Equation~\ref{eqn:Jnu0eta}).

Moderate-dispersion long-slit spectroscopy appears the most promising
technique with which to detect Ly$\alpha$ fluorescence.  From typical
quasar lines-of-sight, there are on average one or more optically-thick
LLS per unit redshift interval at $z >$~2 (Stengler-Larrea \etal\
1995). Hence, a single spectrogram with a slit a few arcmin in length
will intercept several such high-column-density clouds (see
\S\ref{subsec:results}).  Imaging with narrow-band interference filters
(typically $R\approx$~100), although covering a larger solid angle over
a smaller redshift interval, provides less contrast against the sky for
line emission than does spectroscopy. Therefore, narrow-band imaging
cannot achieve the required sensitivity without prohibitively long
integrations. Similarly, the expected surface brightness of Ly$\alpha$
is at least an order of magnitude too faint to be detected in the
deepest broad-band optical imaging, i.e., the Hubble Deep Field
(Williams \etal\ 1996).  Even if sufficient depth could be attained,
since the broad-band covers a large slice of redshift space the
measurement would be confusion limited, as the LLS clouds (expected to
be $\sim$~10\arcsec\ in extent, \S\ref{subsec:results}) would overlap.

In this paper, we present the results of deep Keck~II long-slit
spectroscopy with the Low Resolution Imaging Spectrometer (LRIS, Oke
\etal\ 1995) at moderately-high dispersion in the blue, 150~km~s$^{-1}$
full width half maximum (FWHM).  In \S\ref{sec:obs} we describe the
long-slit spectroscopic observations obtained, detail the data reduction
procedures and determine an upper-limit to $J_{\nu0}$.  We discuss the
cosmological implications in \S\ref{sec:discuss}.

\section{Observations}
\label{sec:obs}

As part of another program, we have been undertaking deep long-slit
spectroscopy on the fields of QSOs whose spectra show damped Ly$\alpha$
absorption systems. One field involved a set-up in the blue with a
high-dispersion grating, and we are able to use the ``blank sky'' in the
long-slit spectrogram for this investigation. The primary targets in
these data are two actively star-forming galaxies at $z=$~2.81 (S2 \& S3
in Warren \& M\o ller 1996) with angular separations of 12\arcsec\ \&
21\arcsec\ from the $z_{\rm em}\approx z_{\rm abs}$ QSO/damped system
PKS0528$-$250, which lies at the same redshift. The coordinates of the
slit center were $\alpha_{2000}=$~05$^{h}$30$^{m}$09\fs 35,
$\delta_{2000}=$~$-$25\arcdeg 03\arcmin 28\farcs 2, and the position
angle was set to 95\fdg 2 East of North so as to intercept both
galaxies.  The spectra of the target objects are to be presented in a
forthcoming paper (Bunker \etal\ 1998).

We obtained the long-slit spectroscopy on the night of 1998 January 20
U.T.\ using LRIS at the f/15 Cassegrain focus of the 10-m Keck~II
Telescope.  The LRIS detector is a Tek 2048$^2$ CCD with 24~\micron\
pixels and an angular scale of 0\farcs 212 per pixel. We read out the
CCD in two-amplifier mode. The observations were obtained using the
900~line~mm$^{-1}$ grating in first order blazed at 5500~\AA, producing
a dispersion of 0.839~\AA~per pixel. The reference arc lamps and
sky-lines have FWHM~$\approx$~3--3.5~pixels, so for objects that fill
the slit, the velocity width of a spectrally unresolved line is
FWHM~$\approx$~150--175~km~s$^{-1}$. The grating was tilted to sample
the wavelength range 4500--6200~\AA, corresponding to Ly$\alpha$ in the
redshift range 2.7~$<z<$~4.1.  The observations used the 1\arcsec\ by
3\arcmin\ long-slit. A total of 5400 s of on-source integration was
obtained, and this was broken into three individual exposures of
duration 1800 s that spanned an airmass range of 1.41--1.53.  The telescope
was dithered by 7\farcs 5 along the slit between each integration to
enable more effective cosmic ray rejection and the elimination of bad
pixels. In the $B$-band the seeing had a FWHM~=~0\farcs 8--1\farcs 1, and
the sky background (no moon) was 22.3~mag~arcsec$^{-2}$.

Spectrophotometric standard stars G191B2B \& HZ44 (Massey \etal\ 1988;
Massey \& Gronwall 1990) were observed at similar airmass 
to determine the spectral shape
of the sensitivity function, but our observations were rendered
non-photometric by intermittent thin cirrus. However, we were able to
boot-strap our spectrophotometry to the narrow-band Ly$\alpha$ line
fluxes from M\o ller \& Warren (1993) of the two $z=$~2.81 star-forming
galaxies on the long slit. As we expect the Ly$\alpha$ emitting patches
to have dimensions much greater than our slit width and to have
approximately uniform surface brightness, the photometric zero-points
were corrected for slit losses using standard star observations with two
different slit widths.

\subsection{Data Reduction}

The initial stages of the data reduction followed standard procedures
and were accomplished using IRAF\footnote{The Image Reduction and
Analysis Facility (IRAF) is distributed by the National Optical
Astronomy Observatories, which are operated by the Association of
Universities for Research in Astronomy, Inc., under cooperative
agreement with the National Science Foundation.}.  All the
calibration and data frames were bias subtracted by fitting a
low-order function to the appropriate over-scan region. Each half of
the CCD was then converted to photo-electrons by correcting for the
gain of the corresponding read-out amplifier. A high
signal-to-noise ratio (SNR) dark frame was then subtracted.

The next step consisted of calibrating the pixel-to-pixel sensitivity
variations of the CCD.  Contemporaneous flat fields were obtained with a
halogen lamp immediately after the science exposures, and these internal
flats were normalized through division by the extracted lamp
spectrum. The SNR for these flat fields was 100--300 per pixel. In order
to remove the residual lower spatial frequency variations in the flat
field due to non-uniform illumination of the slit, we also obtained
spectral dome and twilight-sky flats.  Cosmic rays were identified in
the flat fielded spectra by iteratively calculating the noise in each
column perpendicular to the dispersion axis and identifying discrepant
pixels; those pixels associated with structure sharper than the
point-spread function (PSF) were flagged as cosmic rays.

Wavelength calibration was obtained from arc lamp spectra also taken
immediately after the data frames. To quantify the distortions across
the CCD, we traced the centroid of each arc-line (and the sky-lines in
the data frames) at 10 pixel intervals over the length of the slit (800
pixels).  It was found that the centroids of the lines shifted by up to
3 pixels, comparable to the instrumental resolution.  This distortion
was well fit by a quadratic along the columns (the spatial axis) and a
cubic along the rows (the dispersion direction).  Hence we were able to
form a distortion matrix to map from a pixel space to a wavelength
solution, leaving RMS residuals to the fit of 0.1 pixels (0.08~\AA ).

Rather than doing the sky-subtraction in the usual manner by fitting a
polynomial to each column (which might subtract the extended emission we
are looking for), we first rectified the sky lines using the distortion
matrix.  We then created a high SNR 1D sky spectrum by averaging the
rectified rows. The slit illumination function was determined by
collapsing the two-dimensional (2D) blank sky spectrum dispersion axis,
and this function was used to scale the 1D sky spectrum for row-by-row
sky subtraction.

The individual sky-subtracted, rectified 2D spectra taken at different
dither positions along the slit were then registered and averaged, with
a percentile clipping algorithm used to reject cosmic rays that had
escaped previous cuts.  A mask was used to flag the known bad pixels,
which were excluded from the averaging. The individual 1800~s
integrations were sufficiently long for our data to be background
(rather than read-out noise) limited at all wavelengths. The measured
noise agreed well with the Poissonian statistics of the sky background,
demonstrating that any additional noise introduced by our data reduction
techniques was insignificant.

\subsection{Analysis}

After rectification and sky-subtraction, we searched for extended
Ly$\alpha$ line emission by applying various filters along the spatial
axis of the combined 2D spectrum.  We convolved the 2D spectrum with a
highly-elliptical 2D Gaussian oriented parallel to the slit.  For the
minor axis (along the dispersion direction), the width of the kernel
was set to be comparable to the spectral resolution.  We varied the
length of the major axis to search for signal on any characteristic
scale length between 1\arcsec\ (our seeing disk) and 180\arcsec\ (the
length of the slit). For the anticipated range in $J_{\nu0}$, the
signal per resolution element is not large, so this smoothing is
essential.  We repeated the smoothing with simple boxcar filtering of
various lengths along the slit axis, followed by convolution with a
Gaussian of comparable extent to the seeing disk. This second approach
is optimal if the Ly$\alpha$ clouds do indeed have uniform surface
brightness (Gould \& Weinberg 1996).  For both techniques, visual
inspection revealed no significant extended single-line emission.

The only Ly$\alpha$ line-emission sources detected were the target
galaxies, PKS0528$-$250 S1 \& S2. These have also been observed in
continuum imaging (M\o ller \& Warren 1998), with rest-frame
equivalent widths of $W_{0}=$~88~\AA\ \& 46~\AA\ respectively. This
implies that Ly$\alpha$ arises from an internal UV source (star
formation or AGN activity).  Charlot \& Fall (1993) calculate
$W_{0}=$~80--200~\AA\ for a star forming galaxy, depending on the
initial mass function (IMF). However, selective extinction of this
line through resonant scattering and suppression by dust can greatly
reduce the equivalent width of Ly$\alpha$ in star forming galaxies
(e.g., Steidel \etal\ 1996). In stark contrast, Ly$\alpha$ produced by
external photo-ionization should have a much larger equivalent width,
as the continuum arises mainly from two-photon decays. Hence, the
rest-UV continuum around Ly$\alpha$ provides excellent discrimination
between local OB stars and the diffuse meta-galactic background as the
source of the Lyman continuum photons. Any source with a detectable
continuum can only yield an upper-limit on $J_{\nu0}$, as it will
likely be dominated by local effects.

Our long slit also intercepted a [\ion{O}{2}]~$\lambda$~3727~\AA\
emission line at $z=$~0.424, 61\farcs 5 East of S3, associated with a
$V=$~22.8~mag galaxy.  We detect modest continuum
(W$_0$[\ion{O}{2}]=~100~\AA~$\pm$~10~\AA), and the redshift is confirmed
with [\ion{Ne}{3}]~3869~\AA.  The [\ion{O}{2}] doublet is well resolved,
so our resolution is better than 221~km~s$^{-1}$.

Our search techniques failed to reveal the signature of Ly$\alpha$
fluorescence from optically-thick clouds.  A lack of Ly$\alpha$ emission
constrains the incident UV flux.  In order to test the validity of our
search methods and to place meaningful upper limits on $J_{\nu 0}$, we
examined the recoverability of artificial emission lines added to our
data frames. These synthetic lines simulated clouds of various sizes,
and were convolved with the seeing disk and the instrumental spectral
resolution. For those with shorter spatial extent, less than 5\arcsec ,
we were able to recover 90\% of the simulated spectrally unresolved
lines with total SNR~=~3.  However, for larger artificial clouds, the
recoverability was slightly worse for the same integrated SNR (but lower
surface brightness). This is because slit illumination effects become
more dominant on these larger scale lengths, and the spectra of the
target objects introduce confusion. For SNR~=~3, the recoverability
falls below 60\% on scales exceeding 1\arcmin .

Our typical noise in a 1~$\sq\arcsec$ aperture corresponds to a
Ly$\alpha$ surface brightness of $I_{{\rm Ly\alpha}(-18)}=$~0.4--0.6
(1~$\sigma$). We adopt the convention $I_{{\rm Ly\alpha}(-18)} = I_{{\rm
Ly\alpha}}~/~10^{-18}$~erg~s$^{-1}$~cm$^{-2}$~arcsec$^{-2}$. The quoted
noise is for `clean' regions of the night sky, devoid of strong lines,
and the sensitivity declines slightly at the shorter wavelengths of our
spectral range (4500~\AA~$<\lambda<$~6200~\AA ). These limits are for a
spectral extraction width of 4~pixels (220~km~s$^{-1}$).  This is
optimal for a 1D velocity dispersion of $\sigma_{v}$~=~30~km~s$^{-1}$,
comparable to that measured for Ly$\alpha$ forest clouds at $z\approx
2.3$ (Kim \etal\ 1997). The line widths may in fact be greater due to
coherent velocity structure (shears and bulk flows). For
$\sigma_v>$~20~km~s$^{-1}$, the twin peaks of the Doppler-broadened
resonant profile become resolved, and the SNR falls proportional to
$\sqrt{\sigma_v}$.  We have corrected our limits for $A_{B}=$~0.05~mag
of foreground Galactic extinction (Burstein \& Heiles 1984). The recent
COBE/IRAS dust maps of Schlegel, Finkbeiner \& Davis (1998) indicate a
greater extinction ($A_{B}=$~0.2~mag).
 
Our 3~$\sigma$ upper limits on the Ly$\alpha$ surface brightness as a
function of 1D velocity dispersion and angular cloud size, $\theta$,
projected along our 1\arcsec-wide slit are:
\begin{equation}
I_{{\rm Ly\alpha}(-18)}=
(0.4-0.6)\left(\frac{\theta}{10\arcsec}\right)^{-1/2}
\left(\frac{\sigma_{v}}{30\,{\rm km\,s}^{-1}}\right)^{1/2}\hspace{1.5cm}
(3~\sigma\ {\rm upper~limit})
\end{equation}
for 1\arcsec~$<\theta <$~60\arcsec , 2.7~$<z<$~4.1 and
$\sigma_{v}\gtrsim$~30~km~s$^{-1}$.

\subsection{Upper Limits on $J_{\nu 0}$}
\label{subsec:results}

We use our results to place an upper limit on $J_{\nu 0}$ as a
function of redshift and the size and velocity dispersion of the
clouds. As this is derived from the limiting surface brightness
attained, the constraint on the meta-galactic Lyman continuum flux is
independent of $H_{0}$, $q_{0}$ and $\Lambda_{0}$, and is only subject
to the $(1+z)^{4}$ cosmological surface brightness dimming. The
measurement of $J_{\nu}$ from Ly$\alpha$ fluorescence is immune to the
effects of {\em in situ} dust; the extra-galactic Lyman continuum
photons only penetrate clouds to a depth of $N_{\rm HI}\approx
10^{18}$~cm$^{-2}$ corresponding to $A_{V}\simeq 5\times10^{-4}$~mag
for a Galactic gas-to-dust ratio. Gould \& Weinberg (1996) calculate
that only 4\% of Ly$\alpha$ photons are absorbed as they random-walk
their way out of the nebula from this depth.

An observed Ly$\alpha$
surface brightness $I_{{\rm Ly\alpha}(-18)}$ would correspond to a
meta-galactic background of:
\begin{equation}
J_{\nu0(-21)}=4.76 \; I_{{\rm Ly\alpha}(-18)} \;
\left(\frac{0.5}{\eta_E}\right) \; \left(\frac{\alpha-1}{0.73}\right) \;
\left(\frac{1+z}{4}\right)^{4}
\label{eqn:Jnu0eta}
\end{equation}
where $\eta_E$ is the fraction of energy absorbed from the ionizing
background that is emitted in the form of Ly$\alpha$ photons (Gould \&
Weinberg 1996).  For clouds with column depths exceeding $N_{\rm HI}
\gtrsim 3\times 10^{18}$ cm$^{-2}$, this fraction is robust
($\eta_E\approx$~0.5), but falls to $\eta_E\approx$~0.4 for $N_{\rm
HI}\simeq 10^{18}$ cm$^{-2}$.  Note that our observations constrain
directly the flux of Lyman continuum photons and not $J_{\nu0}$ (the
energy density at the Lyman edge). Conversion to $J_{\nu0}$ assumes a
particular form of the spectrum. Specifically we adopt $J_\nu =
J_{\nu_0}~(\nu/\nu_0)^{-\alpha}$ so that $\int^{\infty}_{\nu_0}
J_\nu~/~(h_{\rm pl}\nu)~d\nu = J_{\nu0}~/~(h_{\rm pl}\alpha)$, where
$h_{\rm pl}$ is Planck's constant. The HM96 spectrum has a slope
$\alpha =$~1.73.

The greatest uncertainty on our upper limits is the size of the clouds.
From a study of the foreground galaxies responsible for
\ion{Mg}{2}~$\lambda\lambda$~2796,~2803~\AA\ QSO absorption systems at
0.2~$<z<$~1.6, Steidel \& Dickinson (1995) estimate the linear extent of
LLS to be $r_{\rm LLS}\sim$~70~$h_{50}^{-1}$~kpc.  For a
randomly-distributed population of spherical clouds of radius $r_{\rm
LLS}$ where the angular size of the clouds greatly exceeds the slit
width (1\arcsec$~\simeq$~7~$h^{-1}_{50}$~kpc at $z\approx$~3), the
expectation value for projected length of intersection $\bar{D}$ along
our spectroscopic slit is:
\begin{equation}
\bar{D}=\frac{\pi r_{\rm LLS}}{2}\pm
\left(\frac{8}{3}-\frac{\pi^{2}}{4}\right)r_{\rm LLS}=(110\pm 14)~\left(
\frac{r_{\rm LLS}}{70~h_{50}^{-1}~{\rm kpc}}\right) ~h_{50}^{-1}~{\rm
kpc}
\label{eqn:slitDist}
\end{equation}
This corresponds to an average angular projection of
$\theta=$~15\arcsec~$\pm$~2\arcsec\ for $q_{0}=$~0.5 and
$\theta=$~10\arcsec~$\pm$~1\arcsec\ for $q_{0}=$~0.1 at $z\approx$~3,
assuming no evolution in the typical size of LLS clouds.  In
calculating our limits on $J_{\nu0}$, we conservatively adopt an
average projection of 10\arcsec\ along the slit, consistent with the
smallest likely size.

The limits we place on $J_{\nu0}$ as a function of redshift are
depicted in Figure~\ref{fig:jnulimits}.  Over the range 2.7~$<z<$~3.1,
a wavelength region for Ly$\alpha$ where the sky spectrum is
relatively featureless, our sensitivity to $J_{\nu 0}$ is
approximately flat. This is due to a combination of LRIS response,
which increases towards the red, and Ly$\alpha$ line energy flux which
decreases as $(1+z)^{-4}$ through cosmological surface brightness
dimming.  In fact, our grating set-up (dictated by the demands of
another project) was slightly red-ward of what may be most desirable
for low-airmass and dark-time observations. The efficiency curve for
LRIS+900~line~mm$^{-1}$ grating would suggest that $z \approx$~2.5 is
the optimal hunting ground (see Figure~\ref{fig:jnulimits}).

Our 3~$\sigma$ upper limit on the meta-galactic flux over the range
2.7~$<z<$~3.1 where we are most sensitive is given by:
\begin{equation}
J_{\nu0(-21)}~=~2.0\ \left(\frac{\theta}{10\arcsec}\right)^{-1/2}
\left(\frac{\sigma_{v}}{30\,{\rm km\,s}^{-1}}\right)^{1/2}\;
\left(\frac{0.5}{\eta_E}\right) \;
\left(\frac{\alpha-1}{0.73}\right)\hspace{1cm} (3~\sigma\ {\rm
upper~limit,}~2.7<z<3.1)
\label{eqn:limz3}
\end{equation}
where $\theta$ is the angular extent of the cloud, and we have assumed a
resonantly-broadened line profile from a cloud with velocity dispersion
$\sigma_v$.  If the optically-thick clouds are as small as 1\arcsec, the
3~$\sigma$ upper limit is only $J_{\nu0(-21)}<$~6.3, which is still a
significant improvement over the previous limit of $J_{\nu0(-21)}<$~200
at 2~$\sigma$ from Lowenthal \etal\ (1990).  At higher redshifts, our
upper limits on $J_{\nu0}$ are:
\begin{equation}
J_{\nu0(-21)}~=~3.0\ \left(\frac{\theta}{10\arcsec}\right)^{-1/2}
\left(\frac{\sigma_{v}}{30\,{\rm km\,s}^{-1}}\right)^{1/2}\;
\left(\frac{0.5}{\eta_E}\right) \;
\left(\frac{\alpha-1}{0.73}\right)\hspace{1cm} (3~\sigma\ {\rm
upper~limit,}~z\approx 3.5)
\label{eqn:limz3.5}
\end{equation}
\begin{equation}
J_{\nu0(-21)}~=~4.0 \left(\frac{\theta}{10\arcsec}\right)^{-1/2}
\left(\frac{\sigma_{v}}{30\,{\rm km\,s}^{-1}}\right)^{1/2}\;
\left(\frac{0.5}{\eta_E}\right) \;
\left(\frac{\alpha-1}{0.73}\right)\hspace{1cm} (3~\sigma\ {\rm
upper~limit,}~z\approx 4.0)
\label{eqn:limz4.0}
\end{equation}

A concern is that we do not know {\em a priori} that our slit actually
intercepts any optically-thick clouds. However, the incidence in QSO
spectra of such clouds with column densities exceeding $N_{\rm HI}$ is:
\begin{equation}
\frac{dn}{dz} = 0.9 \left( \frac{N_{\rm HI}}{10^{18}}\right)^{-0.5}
\left( \frac{1+z}{4}\right)^{1.55}
\end{equation}
 per unit redshift (HM96). Table~1 gives the
likelihood that one or more optically-thick clouds intercept our slit,
where we assume a Poissonian geometric distribution of clouds with a
typical projection of 10\arcsec . Over the redshift interval
2.7~$<z<$~3.1 where our constraints on $J_{\nu0}$ are greatest, there
should be on average $\sim$~1--2 such systems per arcmin surveyed.
Therefore, we can reject the possibility that no 2.7~$<z<$~3.1
optically-thick absorbers lie on our 3\arcmin\ slit at the $>$~97\%
confidence level.

\section{Discussion}
\label{sec:discuss}

Given the null results of our search for line emission, we consider the
cosmological implications of our new upper limits on the ambient Lyman
continuum background at high redshift.
 
\subsection{QSOs and the Proximity Effect}

The measured decrease in the number of Ly$\alpha$-absorbing clouds
induced by the UV radiation field in the neighborhood of a QSO (the
proximity effect) provides an estimate of $J_{\nu0}$ at high redshift,
independent of $H_{0}$.  A recent determination for 1.7~$<z<$~3.8
gives a mean intensity of $J_{\nu0(-21)}$~=~0.3--0.8 (Espey 1993; Lu
\etal\ 1991). These limits are a factor of 2.5--7 below our detection
threshold.  However, the proximity effect is subject to several
systematic uncertainties; $J_{\nu0}$ could be underestimated
if the luminosity of quasars is highly variable, or if most of the
quasars with good spectra (naturally the brightest ones) are magnified
by gravitational lensing, or both. From our results, the global
boosting of the quasar luminosity function cannot exceed a factor of
three.  In addition, the number of clouds near quasars might be
enhanced because of clustering.  If this is not accounted for, it will
lead to an underestimate of the ionizing photon flux.

The QSO luminosity function and its evolution, as derived from
multi-color optical surveys, have been used to estimate the integrated
QSO contribution to the UV background.  A firm lower-limit over
2~$<z<$~3 is $J_{\nu0(-21)}^{\rm QSO}>$~0.23 (Rauch {\em et al.\ }1997).
Accounting for the opacity of H and He associated with intervening
Ly$\alpha$ clouds, Mieksin \& Madau (1993) and HM96 estimate that the
mean intensity of the diffuse radiation field from QSOs increases from
$J_{\nu0(-21)}^{\rm QSO}\approx$~0.01 at the current epoch to 0.6 at
$z=$~2.6--3.0. This is a factor of three below our current upper limits
at high redshift.

On one hand, this estimate of $J_{\nu0}^{\rm QSO}$ is unknown at the
50~per~cent level because of uncertainties in the evolution of the
Ly$\alpha$ forest clouds (Meiksin \& Madau 1993).  On the other, this is
an underestimate because some quasars at these redshifts may be missed
through foreground dust extinction in high-column-density absorbers
(Fall \& Pei 1993). Our upper limit of $J_{\nu0(-21)}<$~2.0 indicates
that at the very most, two out of three QSOs at $z\approx$~3 are missed
through absorption by foreground objects.
This is at the level where we can begin to constrain the models of Fall
\& Pei (1993), who predict that between 10\% and 70\% of $z\sim$~3
QSOs are absent from optical samples.

\subsection{Constraints on High-$z$ Star-Forming Galaxies as the Source
of Lyman Continuum Photons}

It has been suggested that the young, hot OB stars in actively
star-forming galaxies at high redshift may contribute significantly to
the meta-galactic UV background (e.g., Songaila, Cowie \& Lilly 1990).
It is likely that much of the Lyman continuum radiation is absorbed by
\ion{H}{1} in the parent galaxy, i.e., the \ion{H}{2} regions are
ionization and not density bounded.  Our upper limit on $J_{\nu 0}$ at
$z\approx 3$ implies that the number of escaping Lyman continuum photons
from star-forming galaxies can be a factor of no more than two greater
than that contributed by quasars, assuming $J_{\nu0(-21)}^{\rm
QSO}\approx$~0.6 (HM96).

If we take the global star formation density from known high-redshift
galaxies (e.g., Madau \etal\ 1996; Madau 1997), then we can constrain
the escape fraction ($f_{\rm esc}$) of UV photons at wavelengths
short-ward of the Lyman limit for these galaxies.  From the HDF, the
co-moving luminosity density at $\lambda_{\rm rest}\approx$~1500~\AA\
is $\rho({\rm 1500~\AA })=2.1\times
10^{26}~h_{50}$~erg~s$^{-1}$~Hz$^{-1}$~Mpc$^{-3}$ at $z\approx 3$
(Dickinson 1998). For a Salpeter IMF (with
0.1~$M_{\odot}<M_{*}<$~125~$M_{\odot}$), this non-ionizing UV
luminosity translates to an average star formation rate (SFR) per unit
co-moving volume of ${\rho}^{\rm UV}_{\rm SFR}(z\approx 3)=
0.026~h_{50}~M_{\odot}$~yr$^{-1}$~Mpc$^{-3}$, assuming no obscuration.
However, from recent near-infrared observations of nebular emission
lines, Pettini \etal\ (1997, 1998) estimate that the UV continuum at
$z\approx 3$ is extinguished by $A_{1500}\approx$~1--2~mag, implying
that the total star formation rates are a factor of 3--6 higher. This
is broadly consistent with the recent HDF sub-mm detections by
SCUBA/JCMT (Hughes \etal\ 1998) corresponding to ${\rho}^{\rm
FIR}_{\rm SFR}= 0.1~h_{50}~M_{\odot}$~yr$^{-1}$~Mpc$^{-3}$ at
$z\approx 3$. Applying the Calzetti (1997) extinction curve derived
from local star-burst galaxies to the $z\approx 3$ population yields a
similar rest-UV extinction (a factor of 2--6; Calzetti 1998).

A constant star formation rate of 1~$M_{\odot}$~yr$^{-1}$ will produce
$1.44\times 10^{53}$~s$^{-1}$ Lyman continuum photons, from the most
recent version of the spectral evolutionary code of Bruzual \& Charlot
(1993) with the same Salpeter IMF.  Hence, the $z\approx$~3
galaxies produce Lyman continuum photons at a volume-averaged rate of:
\begin{equation}
\dot{n}_{\rm LyC}=2.45\times 10^{-20}~\left(\frac{10^{0.4A_{1500}}}{3}\right)
\left( \frac{1+z}{4}\right)^{3}~h_{50}~{\rm photons~s^{-1}~cm^{-3}}
\end{equation}
where $\dot{n}_{\rm LyC}$ is in physical rather than co-moving
co-ordinates. The mean free path of a Lyman continuum photon, $\Delta
l$, is inferred from Madau (1992), who shows that unit optical depth
is achieved for $\Delta z\approx$~0.2 at $z=3$, corresponding to a
physical length of $\Delta l\approx 40~h^{-1}_{50}$~Mpc (i.e., the
ionizing flux is largely local, and evolutionary effects can be
neglected as Lyman-continuum photons from galaxies at higher redshifts
are either severely absorbed or redshifted to energies below the Lyman
edge). Therefore, the contribution of star-forming galaxies to
$J_{\nu0}$ is given by:
\begin{equation}
J_{\nu0}^{\rm SFR}
=f_{\rm esc}~\frac{h_{\rm pl}\alpha}{4\pi}~\Delta
l~\dot{n}_{\rm LyC}
\end{equation}\
for an ionizing spectrum of the form $J_{\nu}=J_{\nu0}(\nu /
\nu_{0})^{-\alpha}$, which can be expressed as:
\begin{equation}
J_{\nu0(-21)}^{\rm SFR}=0.28~\left( \frac{f_{\rm
esc}}{10\%}\right)~\left(
\frac{\alpha}{1.73}\right)~\left(\frac{\Delta l}{40~h^{-1}_{50}~{\rm
Mpc}}\right)~\left( \frac{{\rho}^{\rm UV}_{\rm SFR}}{0.026~h_{50}}
\right)~\left(\frac{10^{0.4A_{1500}}}{3}\right)
\left( \frac{1+z}{4}\right)^{3}
\end{equation}
Therefore, the dust-corrected SFR per unit co-moving volume at
$z\approx 3$ is equivalent to $J_{\nu 0(-21)}=(3-6)~f_{\rm esc}$, if the
rest-UV is extinguished by factors of 6--3. Hence, our 3~$\sigma$
limit of $J_{\nu 0(-21)}=$~2.0 constrains $f_{\rm esc}<$~25\%--50\%,
assuming QSOs contribute $J_{\nu 0(-21)}^{\rm QSO}=$~0.6 (HM96).
However, Meurer
\etal\ (1997) argue that {\em in situ} dust obscuration leads to a much
greater underestimate of the integrated star formation rate, and that
the rest-UV extinction correction should be a factor of $\approx$~15
rather than 3--6. If this is so, our constraint is significantly
tighter ($f_{\rm esc}<$~10\% at $z\approx$~3), comparable to but still
somewhat above the best limits on the Lyman continuum escape fraction
from galaxies today ($f_{\rm esc}<$~1\%, Deharveng \etal\ 1997).

\subsection{Gunn-Peterson Constraint on the Diffuse IGM}

The absence of an absorption trough on the blue side of the Ly$\alpha$
emission line in the spectra of high-redshift QSOs represents the best
limit on the density of a neutral, smoothly-distributed IGM (Gunn \&
Peterson 1965; Scheuer 1965).  A measurement of the ionizing
meta-galactic flux, when combined with limits on the baryon density from
primordial nucleosynthesis and the Gunn-Peterson optical depth (Jenkins
\& Ostriker 1991; Giallongo, Cristiani \& Trevese 1992) measures what
fraction of gas has collapsed into galaxies and discrete clouds. Our
upper limit on $J_{\nu0}$ requires that baryonic matter in the diffuse
IGM comprises $\Omega_{\rm IGM}<0.016\,h_{50}^{-3}\,\Omega_{\rm crit}$
of the closure density at $z\approx 3$ (from Equation~2 of Songaila,
Cowie \& Lilly 1990). We assume the mean optical depth of the IGM is
$\tau_{z\approx 3}<0.02$ (Steidel \& Sargent 1987). From standard Big
Bang nucleosynthesis (Walker \etal\ 1991), our results imply that the
fraction of baryons residing in the diffuse IGM at $z\approx 3$ is at
most $60\%\,/\,h_{100}$ (where
$h_{100}~=~H_{0}$\,/\,100~km~s$^{-1}$~Mpc$^{-1}$), i.e., no constraint for
low values of H$_0$. We are undertaking deeper integrations which should
improve this upper limit by a factor of three.

\subsection{Decaying Relativistic Neutrinos and the Meta-Galactic UV
Background} 

The recent Super-Kamiokande detection of neutrino oscillation (Fukuda
\etal\ 1998), although providing only a mass difference between neutrino
flavors rather than an absolute measurement of mass, has rekindled
interest in Hot Dark Matter (HDM).  We can use our upper limits on
$J_{\nu 0}$ to constrain any relic decaying neutrino flux (e.g.,
Sciama 1990). According to this theory, relativistic neutrinos make up
the bulk of the critical matter density ($\Omega_{\nu}\approx
0.9\,\Omega_{\rm crit}$), and their decay photons can ionize hydrogen
(i.e., $E_{\gamma}>$~13.6~eV) but not helium. Adopting the values from
Sciama (1995, 1998) for the decay photon energy
($E_{\gamma}\lesssim$~13.8~eV), the neutrino lifetime
($\tau_{\nu}\approx 2\times 10^{23}$\,s) and number density at the
current epoch ($N_{\nu}=112.6\,{\rm cm}^{-3}$) would result in a UV
flux at $z\approx 3$ equivalent to $J_{\nu0(-21)}=4.5$
(Figure~\ref{fig:jnulimits}). This would correspond to a 6~$\sigma$
signal in our spectrum over $2.7<z<3.3$. Our failure to detect such
Ly$\alpha$ emission rules out with high confidence a $\Omega_{\rm
M}=1$ cosmology in which the relativistic decaying neutrinos of Sciama
(1998) form the bulk of dark matter. Furthermore, in a Mixed Dark
Matter (MDM) scenario, an HDM component in the form of
$E_{\gamma}=$~13.8~eV Sciama neutrinos can comprise at most
$\Omega_{\nu}<0.4\Omega_{\rm crit}$.

Our observations constrain the combination of parameters
$(E_{\gamma}-13.6~{\rm eV})~/~(\tau_{\nu}~H_{\circ})$, the excess photon
energy above the Lyman limit available for photo-ionization. If the
neutrinos do account for most of the critical matter density in a flat
Universe (i.e., $\Omega_{\nu}\gtrsim 0.9~\Omega_{\rm crit}$,
$\Omega_{\rm baryon}\lesssim 0.1~\Omega_{\rm crit}$), then
\begin{equation}
(E_{\gamma}-13.6~{\rm eV}) = 0.08~h_{50}\times
\frac{\tau_{\nu}}{2\times10^{23}~{\rm s}}~{\rm eV}\hspace{1cm}
(3~\sigma\ {\rm upper~limit})
\end{equation}
Therefore, the energy of the decay photons can be at most only
0.6~per~cent above the ionization potential of \ion{H}{1}, which would
require much fine-tuning.

\section{Conclusions \& Future Work}

Our moderate-dispersion long-slit spectroscopy with LRIS/Keck has
placed a 3~$\sigma$ upper limit on the ambient UV background at
$z\approx$~3 equivalent to a flux at the Lyman limit of
$J_{\nu0(-21)}<$~2.0, assuming optically-thick clouds with dimensions
$\approx$~10\arcsec. We anticipate even greater sensitivity in the
blue for observations at lower air mass and unaffected by cirrus.
Although this direct limit is almost two orders of magnitude lower
than any previously published, it is still a factor of three above the
expected contribution of known QSOs at these epochs.  However, it does
enable us to say that the completeness of optical QSO catalogs is
better than 30\%, and that the contribution to
$J_{\nu0}$($z\approx$~3) from star-forming galaxies cannot exceed
twice that from known quasars. We constrain the escape fraction of
Lyman continuum photons from star-forming galaxies at these redshifts
to be $f_{\rm esc}<$~10\%--50\% (assuming extinction corrections of
15--3 at $\lambda_{\rm rest} \approx$~1500~\AA). Although our
relatively short (5400~s) integration in this pilot study is
insufficient to reach the lower limit on the quasar contribution to
$J_{\nu 0}$, we are able to exclude the possibility that relic
neutrinos comprise the bulk of dark matter, as we would detect
Ly$\alpha$ fluorescence from the flux of ionizing decay photons.

The size of optically-thick QSO absorption systems is largely
unconstrained (\S\ref{subsec:results}), and their morphology is entirely
unknown because quasar sight-lines yield only one-dimensional
information. Some work has been done using two lines-of-sight provided,
for example, by a pair of lensed quasar images (e.g., Dinshaw \etal\
1994).  The single example of multiple sight-lines to a damped system
indicates a size greater than $16\,h^{-1}_{50}$\,kpc (Briggs \etal\
1989).  Without strong constraints on the sizes of the
high-column-density systems, it is impossible to differentiate between
the scenario where outer galactic halos are directly responsible for the
absorption or whether the culprit is more extended diffuse gas. This may
be associated with a cluster, possibly with a fragmented, filamentary
topology (as suggested by recent SPH simulations, e.g., Katz \etal\
1996).  A detection of fluorescent Ly$\alpha$ would reveal the size and
morphology of the optically-thick clouds, which is a primary clue to
understanding their origin and relation to the galaxies which are
forming at these epochs.

The phenomenal light-grasp of Keck plus the new insights afforded by
Gould \& Weinberg (1996) make it possible to design an experiment with
an extremely high likelihood of detecting fluorescent Ly$\alpha$
emission from high-column Lyman limit absorbers. This will enable the
first direct measurement of the size of these high-redshift \ion{H}{1}
clouds, and the determination of $J_{\nu0}$ at $z\approx$~2.5--3.  This
initial study has shown that, for a dark-time integration of
$\sim$~20~hours in good conditions, the sensitivity of LRIS/Keck should
be sufficient to detect the Lyman continuum background at
$>$~5~$\sigma$, even if the meta-galactic flux has no major contribution
other than the known high-redshift quasars.  We are embarking on an
extensive program of LRIS/Keck observations to obtain long-slit spectra
of several hours. By using a filter of width
$\Delta\lambda/\lambda\approx$~10\%, we can obtain several parallel
long-slits on the same spectrogram, thereby covering a large solid angle
while concentrating on those redshifts ($z\approx $~2.5) where our
sensitivity to $J_{\nu}$ is greatest (Figure~\ref{fig:probz}).  For the
first time we will set strict quantitative limits on the inventory of
sources of UV radiation at high redshift.

\acknowledgments We are indebted to Hy Spinrad and Steve Warren for
allowing us to use their spectroscopic data for this blank-sky
search. Our thanks to Daniel Stern, Arjun Dey and Adam Stanford for
their help during the observing run. The software for rejecting cosmic
ray strikes was written by Mark Dickinson. Douglas Finkbeiner kindly
provided us with the extinction from the COBE/IRAS maps.  We gratefully
acknowledge invaluable discussions with Joe Silk, Nick Gnedin, Marc
Davis, Neal Katz, Mark Lacy \& Chuck Steidel.  Palle M\o ller graciously
provided us with imaging of this field. Our thanks to Daniel Stern,
Michael Liu and Leonidas Moustakas for comments on an early version of
this manuscript. We appreciate the constructive suggestions of our
referee, Andrew Gould. The data presented herein were obtained at the
W. M.\ Keck Observatory, which is operated as a scientific partnership
among the California Institute of Technology, the University of
California and the National Aeronautics and Space Administration.  The
Observatory was made possible by the generous financial support of the
W. M.\ Keck Foundation.  We received excellent support while observing
at Keck, and we are grateful to Bob Goodrich, Fred Chaffee, Tom Bida,
Terry Stickel \& David Sprayberry. A.J.B.\ acknowledges a NICMOS
postdoctoral research fellowship. F.R.M.\ would like to acknowledge
support from the HST NASA grant \#AR-07523.01. J.R.G.\ is supported by
the Packard Foundation.

\clearpage

\begin{deluxetable}{rcc}
\tablewidth{33pc} \tablecaption{Upper Limits and Detection Statistics
\label{tab:ProbIntercep}}
\tablehead{
\colhead{} & \multicolumn{2}{c}{$N_{\rm HI}$~/~cm$^{-2}$} \nl
& $>10^{18}$ & $>3\times 10^{18}$}
\startdata
$\eta_{E}$ & 0.4 & 0.5 \nl
$J_{\nu0(-21)}$ 3~$\sigma$ limit & 2.5 & 2.0 \nl
\#~of~clouds~/~arcmin & 2.0 & 1.2 \nl
P($n_{\rm cloud}\ge1$) & 99.8\% & 97.3\% \nl
\tablecomments{For the redshift interval 2.7~$<z<$~3.1 where
our limits on $J_{\nu 0}$ are most stringent.}
\enddata
\end{deluxetable}

\clearpage
\figcaption[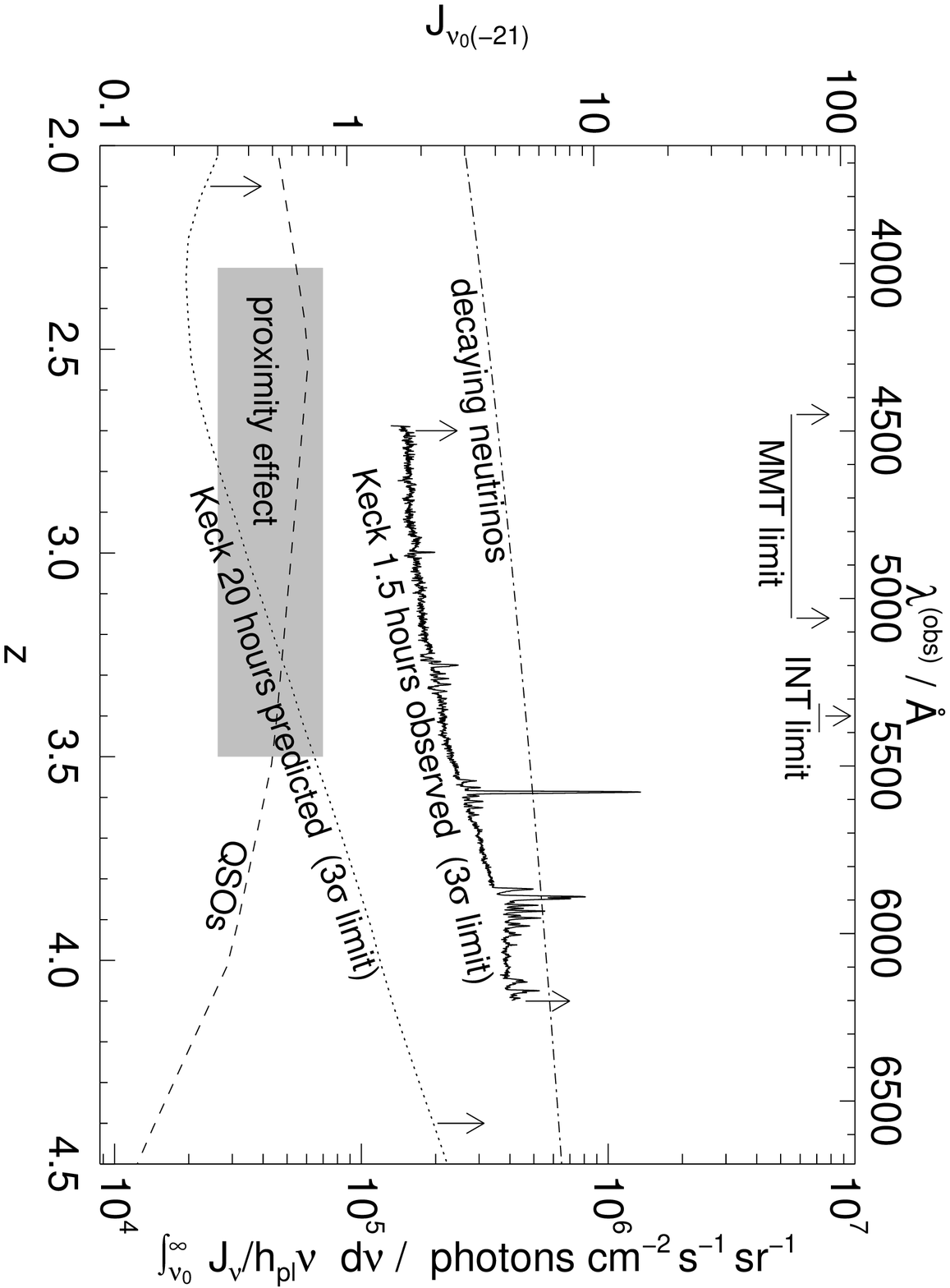]{Our 3~$\sigma$ upper-limits on the flux
density at the Lyman edge,
$J_{\nu0(-21)}=J_{\nu0}~/~10^{-21}$~erg~s$^{-1}$~cm$^{-2}$~Hz$^{-1}$~sr$^{-1}$,
as a function of redshift (solid line). We assume a meta-galactic UV
spectrum of the form $J(\nu)\propto \nu^{-\alpha}$ with $\alpha=$~1.73
(HM96). The right axis shows the flux of Lyman continuum photons
($\int^{\infty}_{\nu_0} J_\nu/h_{\rm pl}\nu~d\nu$), which is measured
independent of $\alpha$. The region above the heavy solid curve is
excluded at the 90\% confidence level. Spikes of reduced sensitivity
occur at redshifts where Ly$\alpha$ coincides with the sky lines in
our 1.5~hour spectrogram. These limits assume that a long 1\arcsec\
slit covers $\sim$~10~$\sq$\arcsec\ of an optically-thick cloud
($N_{\rm HI} \gtrsim 3\times 10^{18}$~cm$^{-2}$) and that the cloud
velocity dispersion is $\sigma_v \lesssim$~30~km~s$^{-1}$.  Also shown
are the previous 3~$\sigma$ upper-limits from the INT
(Mart\'{\i}nez-Gonz\'{a}lez 1995) and the MMT (Lowenthal
\etal\ 1990), converted to the same assumed projected cloud size.  We
plot our anticipated sensitivity for a 20~hour exposure with
LRIS+900~line~mm$^{-1}$ at low airmass in dark time (dotted curved).
The shaded region is the lower-limit on $J_{\nu0}$ from the proximity
effect (Espey 1993). The estimated contribution to $J_{\nu0}$ from the
luminosity function of known high-$z$ QSOs is also shown as the dashed
line (HM96). The equivalent Lyman continuum flux from decaying relic
neutrinos is plotted as a dot-dash line, adopting the parameters of
Sciama (1998) and assuming these neutrinos form the bulk of the dark
matter in an $\Omega_{\rm M}=1$ Universe. Such a scenario is strongly
ruled out by the null results of our survey.
\label{fig:jnulimits}}

\clearpage
\begin{figure}[h]
\plotone{figure1.ps}
\end{figure}

\clearpage
\figcaption[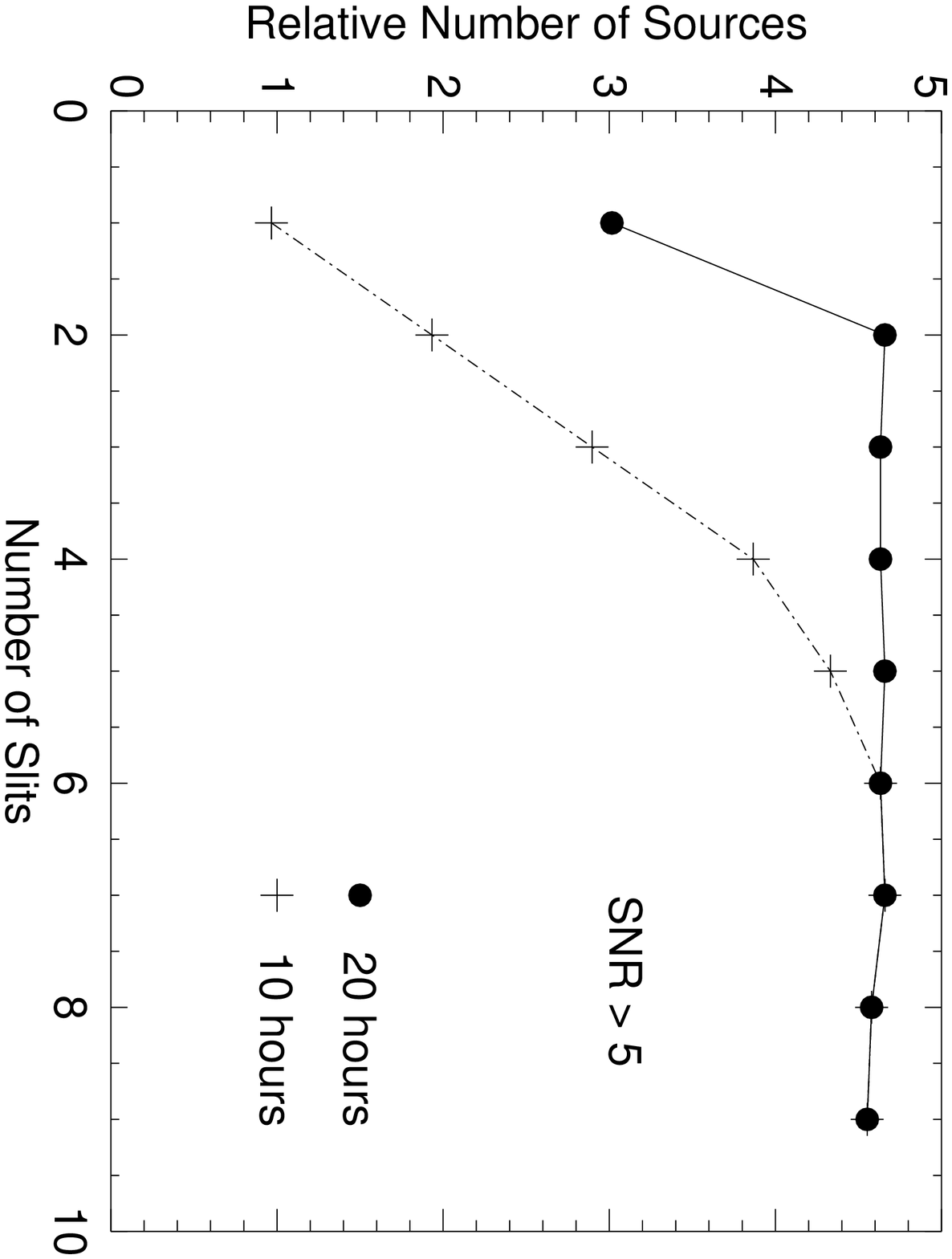]{The number of sources intercepted above a
	given SNR threshold (SNR~$>$~5) in a 10--20~hour spectrum as a
	function of the number of slits. We perform this calculation for
	the LRIS spectrograph on Keck with the 900~line~mm$^{-1}$
	grating set to $\lambda_{\rm cent}=$~4047~\AA .  Given the
	redshift distribution of Ly$\alpha$ absorbers and our limits on
	$J_{\nu0}$ as a function of wavelength
	(Figure~\ref{fig:jnulimits}), we are most sensitive to clouds
	over a relatively limited range of redshifts. This means that
	pixels in the detector array are better used in the spatial
	dimension rather than the spectral dimension.  Having multiple
	slits trades wavelength range for area covered. Since LRIS is
	equipped with a slit mask mechanism we can insert several
	parallel long slits of length $\approx$~7\arcmin\ at the focal
	plane. A custom narrow band filter ($\Delta\lambda/\lambda
	\approx$~10\%) would be used to prevent spectral overlap.  This
	figure shows the increase in the efficiency by using this
	multi-slit Ly$\alpha$ search technique.  We conservatively
	assume that the number of clouds per unit redshift interval
	scales as (1~$+z$), which is correct for damped Lyman alpha
	systems (Lanzetta, Wolfe \& Turnshek 1995).  For lower-column
	systems, $dn/dz \propto (1+z)^{\gamma}$, where
	$\gamma\approx$~1.5 (Stengler-Larrea \etal\ 1995).  Adoption of
	this steeper distribution would favor a multi-slit search even
	more.
\label{fig:probz}}

\clearpage
\begin{figure}[h]
\plotone{figure2.ps}
\end{figure}

\end{document}